# Predicting the DNA Conductance using Deep Feed Forward Neural Network Model

Abhishek Aggarwal[1,#], Vinayak[2,#], Saientan Bag[1], Chiranjib Bhattacharyya[3], Umesh V. Waghmare[4], and Prabal K Maiti[1, *]

[1] Center for Condensed Matter Theory, Department of Physics, Indian Institute of Science, Bangalore 560012

[2] Undergraduate program, Indian Institute of Science, Bangalore 560012

[3] Department of Computer Science and Automation, Indian Institute of Science, Bangalore 560012

[4] Theoretical Sciences Unit, Jawaharlal Nehru Center for Advanced Scientific Research, Jakkur P.O., Bangalore 560064, India

*Abstract*— **Double-stranded DNA (dsDNA) has been established as an efficient medium for charge migration, bringing it to the forefront of the field of molecular electronics as well as biological research. The charge migration rate is controlled by the electronic couplings between the two nucleobases of DNA/RNA. These electronic couplings strongly depend on the intermolecular geometry and orientation. Estimating these electronic couplings for all the possible relative geometries of molecules using the computationally demanding first-principles calculations requires a lot of time as well as computation resources. In this article, we present a Machine Learning (ML) based model to calculate the electronic coupling between any two bases of dsDNA/dsRNA of any length and sequence and bypass the computationally expensive first-principles calculations. Using the Coulomb matrix representation which encodes the atomic identities and coordinates of the DNA base pairs to prepare the input dataset, we train a feedforward neural network model. Our NN model can predict the electronic couplings between dsDNA base pairs with any structural orientation with a MAE of less than 0.014 eV. We further use the NN predicted electronic coupling values to compute the dsDNA/dsRNA conductance.**

[#] Equal Contribution
*email: maiti@iisc.ac.in, phone: 091-80-2293-2865

## Section 1. Introduction

In the past few years, the availability of large datasets along with constant improvements in algorithms and the unprecedented growth in computational power has driven machine learning (ML) to the forefront of research in almost every scientific field[1-8]. ML has been used in numerous theoretical works with the purpose of bypassing the computationally demanding Density Functional Theory (DFT) calculations[9-15]. Replacing the traditional techniques, ML has already been used to develop atomistic force field parameters with first-principles accuracy[10], to predict the models for catalysis[16] as well as to identify new materials for molecular electronics and predict their electrical properties[1, 17-21]. From implementation in speech recognition and image classification[22] to prediction of pharmaceutical properties of molecular compounds and targets for drug discovery[5, 23, 24], ML methods are being used in every aspect of research. Recently, few studies were aimed at predicting the charge migration properties of different materials using ML methods[1, 12, 17, 25-29]. Lederar et al.[17] predicted the electronic couplings between pentacene molecules in a crystal by feeding their relative orientation vectors as input to Kernel Ridge Regression (KRR) algorithm. Korol et al.[1] used the DNA sequence information as input to NN architecture to predict its conductance. Caylak et al.[25] used a Feed Forward Neural Network (FFNN) optimized using genetic algorithm to predict the electronic coupling for the amorphous morphology of Alq3. Bag et al.[29] recently predicted the electronic coupling between two guanine bases using their coarse-grained representation as feature vectors. Similarly, Wang et al.[12] predicted the electronic coupling values of ethylene dimers by providing the input in the form of Coulomb matrix to Kernel Ridge Regression (KRR) algorithm. However, the electronic coupling between two molecules of different types has not been predicted yet in any of the works.

Charge transport through DNA, the genetic material of life, has attracted a lot of attention in the last few years owing to its great biological as well as physical significance[30-35]. Long-range charge migration in DNA plays a crucial role in various biological phenomenon such as oxidative damage in DNA, DNA repair signaling, and mutation[31, 32, 36]. It is established that for small-length dsDNA, charge transport is dominated by tunneling mechanism[33, 35, 37], while thermally-induced charge hopping mechanism takes over for long double-stranded DNA (dsDNA)[33, 38-40]. Experimentally, the electrical conductance of a dsDNA/dsRNA molecule is typically measured at room temperature and in the presence of a solvent[41-43]. Its realistic estimation thus needs a combination of large-scale molecular dynamics simulation with access to quantum probabilities of electron transfer through the molecule, which is computationally intensive[33, 39, 40, 44]. Many theoretical studies have been reported on dsDNA/dsRNA to understand their conductive properties[37-40, 42-51]. However, because of the computational constraints, these studies are limited to only short dsDNA/dsRNA sequences[39, 40, 44]. For example, to explore all the possible sequences of 12 bp long dsDNA, $12^4$ sequences will need to be studied which is an implausible task if done using DFT calculations. Therefore, a faster tool which can be used to predict the conductance of a given dsDNA sequence of any length is required.

Since the electronic coupling is the most time-consuming part of a charge transport calculation, in this work we use ML to bypass the DFT calculations and predict the transfer integrals between any two dsDNA/dsRNA nucleobases. The ML algorithms proposed earlier for modelling charge transport can predict transfer integrals between same entities[1, 12, 17]. However, as there are 5 types of nucleobases in dsDNA/dsRNA, the charge can transfer between different nucleobases[39, 40, 44]. Here, we present an NN-based model that predicts the quantum probability of electron transport across any base-pair in these molecules knowing only the positions of atoms which are directly available in an MD simulation. Capturing the dependence of electron transport on the relative orientation and chemistry of bases, our method can be used within any MD simulation of these molecules to estimate their electrical conductance. We then use our methodology to gain insight to various biological phenomena where we verify the validity of our approach by predicting the differences between conductance of dsDNA and dsRNA and from the dependence of DNA conductance on its length. We further explore the applications of our ML model by detection of a single mismatch in dsDNA sequence using NN predicted electrical properties. Since the charge transport in DNA occurs through base pairs in the thermally-induced hopping regime[39, 40, 44, 52], the ability to predict the electronic coupling between any two nucleobases opens up the possibility to predict the conductance of a dsDNA of any length or any sequence.

## Section 2. Methodology and Results

**Section 2.1 MD Simulation**

The initial structures of all the dsDNA/dsRNA were built using the NAB module of AmberTools18[53], which were then solvated in a TIP3P[54] water box using the xLEAP module of AmberTools18[53]. Appropriate number of $Na^+$ counterions were added to the system to neutralize the negatively charged phosphate backbone. A combination of ff99bsc0 and OL3[55, 56] force fields and a combination of ff99bsc0 and OL15[57, 58] force fields was used in the dsRNA and dsDNA simulations, respectively. This system is then simulated using MD with the protocol which is described in detail in our earlier publications[39, 40, 44, 59, 60]. For the data analysis and the visualization of the systems, we have used Visual Molecular Dynamics (VMD)[61] package extensively.

To construct an input dataset that contains information on all the base pair combinations of dsDNA/dsRNA, we simulate a variety of dsDNA/dsRNA sequences of different lengths as listed in Table 1. We simulate d-$(CG)_n$ for both dsDNA and dsRNA as well as dsRNA of sequence d-$(CGCGA_nU_nCGCG)$, where n varies from 2 to 6. The inputs from these sequences contribute to the length variation in input dataset. To include inputs for other base pair combinations, we simulate different 12 bp dsDNA and dsRNA sequences, namely, d-$(AA)_6$, d-$(AU)_6$, d-$(CG)_6$ and d-$(CC)_6$ for both dsDNA as well as dsRNA. We then choose 100 morphologies of dsDNA/dsRNA from the last 2 ns of the 100 ns MD simulated dsDNA/dsRNA trajectories and compute the transfer integrals for all the base pairs. We also simulate Drew-Dickerson dsDNA sequence (d-(CGCGAATTCGCG)) with and without a single A-G mismatch[62] as shown in Fig. 5a) to explore the capabilities of our NN model.

| S. No. | dsDNA Sequence | dsRNA Sequence | Length of Sequence |
|---|---|---|---|
| 1. | d-(CGCGAATTCGCG) | d-(CGCGAAUUCGCG) | 12 bp |
| 2. | d-(CGCGCGCGCGCG) | d-(CGCGCGCGCGCG) | 12 bp |
| 3. | d-(AAAAAAAAAAAA) | d-(AAAAAAAAAAAA) | 12 bp |
| 4. | d-(ATATATATATAT) | d-(AUAUAUAUAUAU) | 12 bp |
| 5. | d-(CGCGAATTCGCG) | d-(CGCGAAUUCGCG) | 12 bp |

| 6. | d-(CG)$_n$ | d-(CG)$_n$ | 4 bp to 12 bp |
| 7. | – | d-(CGCGA$_n$U$_n$CGCG) | 12 bp to 20 bp |

Table 1 List of dsDNA and dsRNA simulated and used for electronic couplings calculations and training of the NN model.

### Section 2.2 Hopping charge transport mechanism

In the Semi-Classical Marcus Hush formalism[63], the charge transport through a molecule is described as incoherent hopping motion of charge carriers between available sites[39, 40]. Several theoretical and experimental investigations have demonstrated that the charge transport in nucleic acids is mediated by stacked nucleobases through strong π-π interactions[39, 40, 52, 64]. Hence, we consider the nucleobases as charge hopping sites and replace the backbone atoms with hydrogen atoms in further calculations and optimizations as shown in Fig. 1a.

In Marcus-Hush formalism, the charge hopping rate $\omega_{ik}$ from charge hopping site, $i$, to the hopping site, $k$, is given by

$$\omega_{ik} = \frac{2\pi |J_{ik}|^2}{h} \sqrt{\frac{\pi}{\lambda k_B T}} \exp\left[-\frac{(\Delta G_{ik} - \lambda)^2}{4\lambda k_B T}\right] \quad (1)$$

Where $J_{ik}$ is the electronic coupling, also called transfer integral, defined as:

$$J_{ik} = <\phi^i|H|\phi^k> \quad (2)$$

Here $\phi^i$ and $\phi^k$ are diabatic wave functions localized on the sites $i$ and $k$, respectively. H is the Hamiltonian for the two-site system between which the charge transfer takes place. λ is the reorganization energy. $\Delta G_{ik}$ is the free energy difference between the two sites, $h$ is the Plank's constant, $k_B$ is the Boltzmann constant, and $T$ is the absolute temperature.

The calculation of transfer integrals and reorganization energies were performed using density functional theory (DFT) which have been carried out using Gaussian 09[65] software with M062X/6-31g(d) level of theory. VOTCA-CTP[66] software package is used to calculate the transfer integral values for all possible base pairs.

## Section 2.3 Kinetic Monte Carlo (KMC) Calculations

Once all the charge hopping rates are obtained for all possible base pairs, Kinetic Monte Carlo (KMC) method[39, 40, 66] is used to solve the probability master equation and obtain the charge dynamics to calculate the V-I characteristics of the DNA/RNA molecule. This is done by assigning a unit charge at a random hopping site, $i$, at initial time $t = 0$. The total escape rate for the charge at site $i$ to all the possible neighbouring sites ($N$) is given as

$$\omega_i = \Sigma_{k=1}^{N} \omega_{ik} \tag{3}$$

The waiting time ($\tau$) is then calculated using $\omega_i$ and a uniform random number ($r_1$) between 0 and 1, as

$$\tau = -\omega_i^{-1} \ln(r_1) \tag{4}$$

The site to which the charge will hop to is chosen as the one for which the expression $\frac{\Sigma_k \omega_{ik}}{\omega_i}$ is largest and is $\leq r_2$, where $r_2$ is another uniform random number between 0 and 1. The above condition is to ensure that the site $k$ is chosen with probability $\frac{\omega_{ik}}{\omega}$. After this, we update the position of the charge (to $k$) as well as the total time of the system (as $t = t + \tau$) and repeat the above process which provides the probability of finding the charge at each site. The total current can then be found using the following formula

$$I_{bp} = -e \left[ \Sigma_i (P_{b_1} \omega_{b_1 i} - P_i \omega_{ib_1}) + \Sigma_i (P_{b_2} \omega_{b_2 i} - P_i \omega_{ib_2}) \right] \tag{5}$$

Here, e is the unit electric charge, $i$ stands for all the possible hopping sites which are in the direction of flow of current, b₁ and b₂ are the base stacks of base pair bp. Hence mean current is average over all base pairs, $I = <I_{bp}>$.

## Section 2.4 Training a Deep Neural Network (NN)

To correctly predict the electronic couplings between different dsDNA base pairs, a descriptor is required which can capture the constitution, the relative positions, and the relative orientations of the two nucleobases. We use Coulomb matrix ($M_{ij}$) representation[25, 67, 68] to construct the input feature vector for a base pair which encapsulates these structural features. Since different dsDNA/dsRNA nucleobases have different number of atoms, we take the default matrix size corresponding to the base pair combination with the largest number of atoms (GG base pair), $k$, to keep the dimension of feature vectors same for all base pair combinations. For base pair combinations with lesser number of atoms, we add zero padding to the Coulomb matrix i.e. assign zero to the empty matrix elements. Hence, for a given base pair combination, the matrix element, $m_{ij}$, of the corresponding Coulomb matrix is defined as[25, 67, 68]:

$$m_{ij} = \begin{cases} \frac{1}{2} Z_i^{2.4} & \text{, for } i = j \text{ and } i,j \leq N \\ \frac{Z_i Z_j}{\|R_i - R_j\|} & \text{, for } i \neq j \text{ and } i,j \leq N \\ 0 & \text{, for } N < i,j \leq k \end{cases}$$

where, $Z_{i(j)}$ is the nuclear charge of the atom $i(j)$, $R_{i(j)}$ are the atomic coordinates of the atom $i(j)$ and $N$ is the number of atoms in a base pair. Since the hopping rate between two molecules depends on the square of the transfer integral (eq. 1), we scale the transfer integral as $F = log_{10}[|J_{ij}|^2]$. $F$ is further normalized to have a mean of 0 and a variance of 1. The final Coulomb matrix is constituted of four block square matrices, where the diagonal matrices represent the self-interaction of the nucleobases while the off-diagonal matrices represent the inter-base interaction terms as shown in Fig. 1b. This matrix is then further pre-processed before being fed to the NN algorithm as input vector as depicted in Fig. 1c. Since the Coulomb matrix is symmetric[25], only the upper triangle elements of the matrix are kept for the NN algorithm. This upper triangular matrix is then further converted to a single-column vector input for the FFNN algorithm. We use scikit-learn[69] Python package for all the NN calculations.

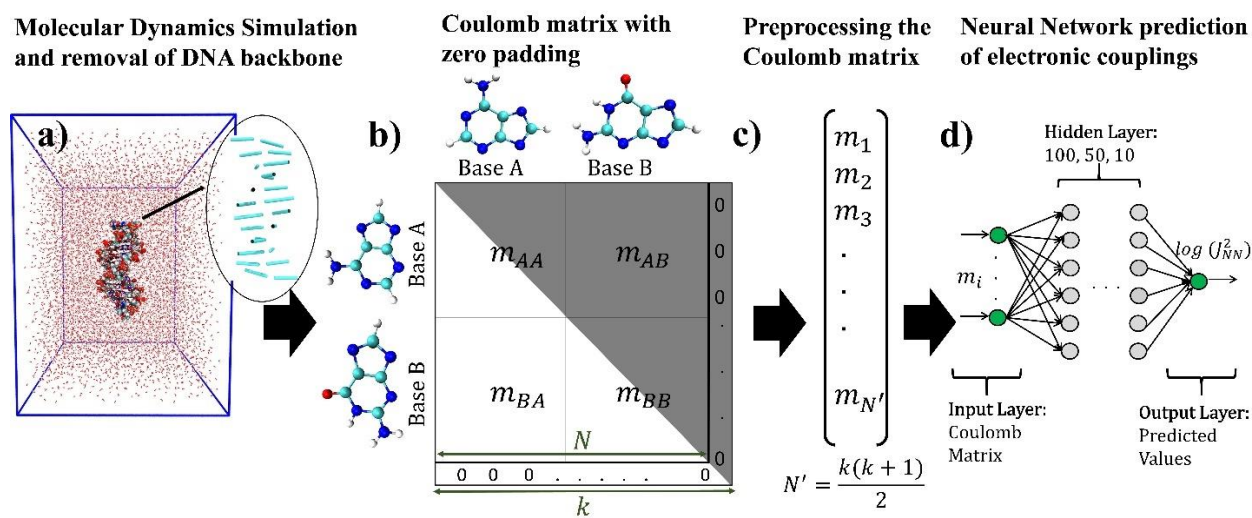

Fig 1 Schematic diagram to represent the process of data flow from molecular dynamics simulations to neural network input. a) Atomic coordinates of different base pairs are collected from the 100 MD-simulated-morphologies. b) For each base pair combination, Coulomb matrix is prepared with zero padding. c) The Coulomb matrix is then pre-processed using feature selection (upper triangle) and data scaling to have a mean of 0 and variance of 1 to remove the redundant information. d) The one-dimensional Coulomb matrix of dimension $N' = \frac{k(k+1)}{2}$ is then fed to a NN system of 3 hidden layers (100,50,10), which predicts the transfer integral values.

## Section 3. Results

### Section 3.1 Prediction of the Trained Neural Network (NN)

The input dataset obtained using all the dsDNA/dsRNA sequences simulated in this work leads to a big dataset of ~60,000 entries. We use a feedforward neural network (FFNN) which consists of multiple hidden layers, each layer consisting of a finite set of neurons connected to maintain unidirectional flow of information[25]. The training of the neural network is done using mini-batch gradient descent using backpropagation algorithm. The number of layers and the number of nodes in each layer (usually referred to as the hyper-parameters) were optimized using grid search algorithm coupled with k-fold cross validation (with k=5). The search is done among the ranges as depicted in Table S1 of the Supplementary Information. For our dataset, NN having 3 hidden layers with 100, 50, 10 nodes using the training data with "sigmoid" activation function gives the best accuracy. The efficiency of the training was evaluated by comparing the DFT calculated electronic couplings and the NN predicted electronic couplings on the test set. We use Mean Absolute Error (MAE) to characterize the model's accuracy and precision which is defined as, $MAE = \frac{\sum_M |J_{NN} - J_{DFT}|}{M}$, where $M$ is the total number of predicted values.

Fig. 2a shows that our FFNN model predicts the test set as well as the validation set with an MAE of 0.011 eV which is comparable to the one achieved by Lederer *et. al*[17]. For small molecule such as ethylene, Wang et al.[12] achieved an MAE of 0.004 eV, while Çaylak et al.[25] achieved an MAE of 0.55 for $log_{10}[|J_{ij}|^2]$ for Alq$^3$ molecules. We show in Fig. 2b that our achieved MAE of 0.011 eV is sufficient for prediction of accurate charge transport properties of dsDNA. We use a random assignment ratio of 80:10:10 for training, validation, and testing purposes, respectively. We then compute the I-V characteristics for 25 different morphologies of Drew-Dickerson dsDNA sequence chosen randomly from the 100 ns long MD simulation using both NN predicted electronic coupling values as well as DFT calculated electronic coupling values (Fig. 2b). Clearly, the trend of I-V curve as well as the order of magnitude is same for both the cases and the actual values of the current predicted using NN predicted couplings fall well within the error bar range of the current computed using DFT calculated electronic coupling values. This marks the validity of our approach. Fig. 2c shows the training scores for different sizes of training dataset. A training score of 1 refers to the perfect model which predicts accurately. The small difference between the training score and the cross-validation score after a training dataset size of ~15,000 shows that even a dataset of less than 15% entries of the dataset will also predict the transfer integral values with a similar accuracy.

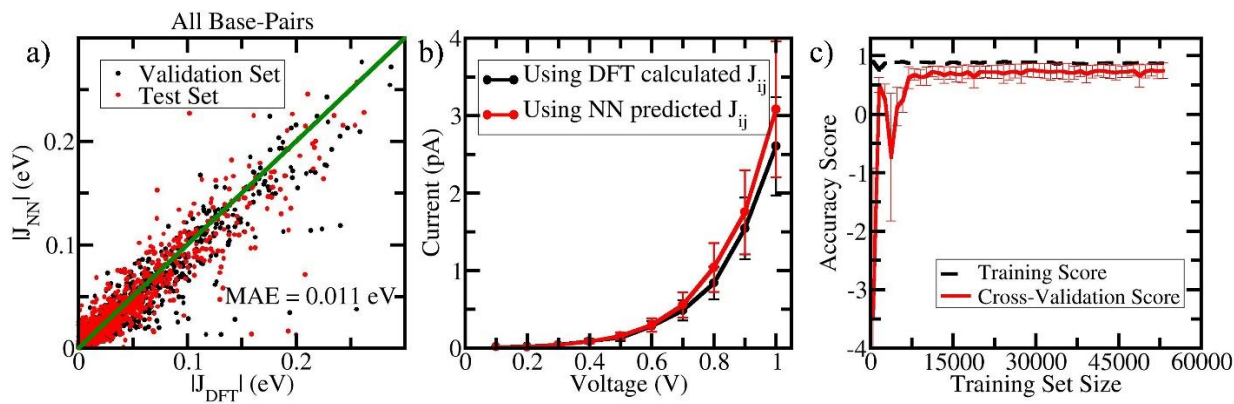

Fig. 2 NN prediction results. a) Correlation between the NN predicted values of the electronic coupling vs the DFT calculated electronic couplings. The green line represents the ideal case when NN predicts an electronic coupling matches exactly with the DFT predicted value. When given unknown data (not used for training) as input, the NN predicts the electronic coupling with an accuracy of 0.011 eV (MAE) of the actual DFT calculated value. b) I-V characteristics calculated for 25 different frames of Drew-Dickerson dsDNA sequence using both DFT calculated electronic couplings as well as NN predicted electronic coupling values. The current calculated using NN predicted couplings matches well with that of DFT calculated couplings. c) The accuracy score for the training and cross-validation for the model. An accuracy of 1 refers to the perfect model which predicts accurately. The small difference between the training score and the validation score after ~15,000 training set size shows that the smaller training datasets will also predict the electronic couplings accurately.

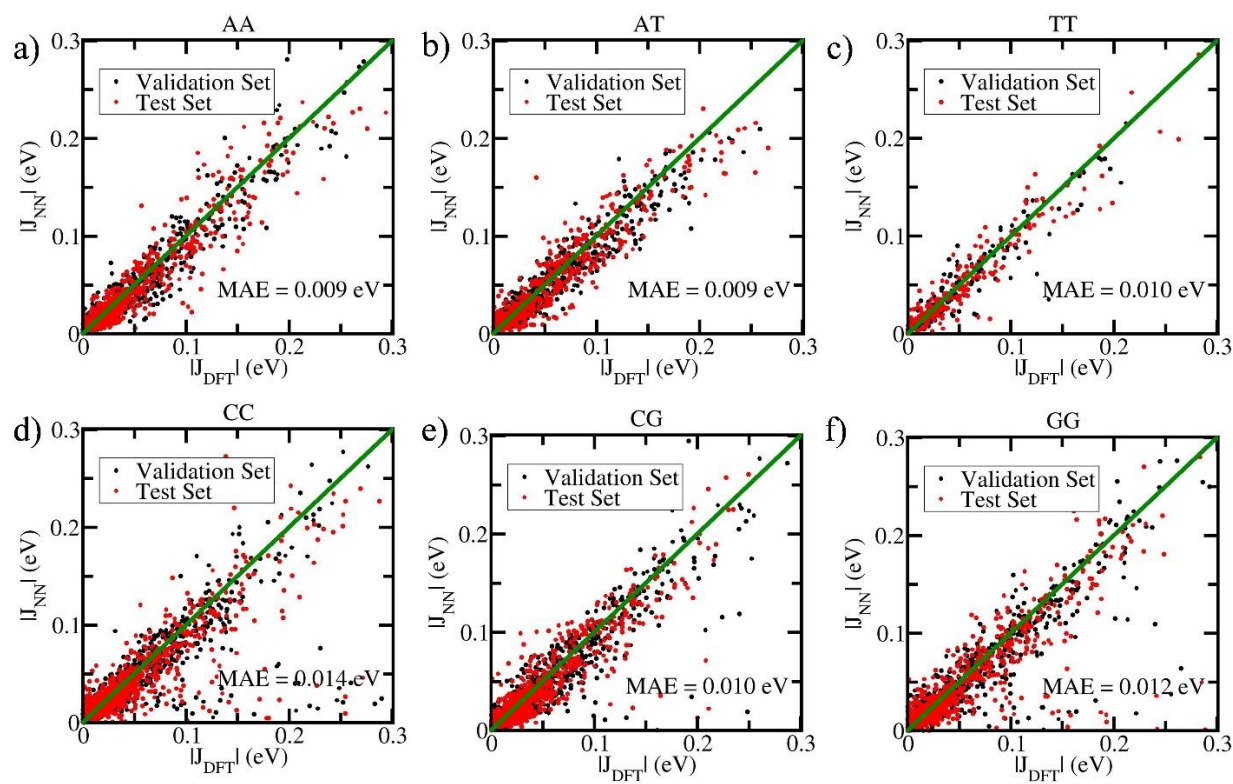

Fig. 3 NN prediction results for different base-pair combinations. Correlation between the NN predicted values of the electronic coupling vs the DFT calculated electronic couplings for a) AA, b) AT, c) TT, d) CC, e) CG and f) GG base-pairs. The MAE for all the base pairs is less than 0.014 eV. The green lines represent the ideal case of NN predicted electronic couplings to be equal to DFT calculated values.

We then predicted the electronic coupling values for all the possible base pair combinations which are shown in Fig. 3 and Fig. S1 of the Supplementary Information. Clearly, our NN model can predict the electronic couplings for each base pair coupling with a MAE less than 0.014 eV. Notably as shown in Fig. S1, for the base pair combinations which have very few number of test set data points (for example AG and GU), our NN model predicts the electronic coupling values with a similar MAE marking the generality and robustness of our approach.

**Section 3.2 Applications of the NN model**

To test the applicability of our ML model, we first reproduce known electrical properties of B-form dsDNA and A-form dsRNA[33]. We calculate the electrical conductance of dsDNA and dsRNA for 2500 morphologies taken from the last 50 ns simulated trajectory using the ML predicted transfer integral values. Figure 4a shows the average V-I characteristics for B-form Drew Dickerson dsDNA and corresponding A-form dsRNA computed using DFT calculated electronic couplings for 100 different morphologies taken from MD simulation as well as using NN predicted electronic couplings for 2500 different morphologies taken from the 100 ns long MD simulation trajectory. Clearly, our NN model reproduces the I-V characteristics for B-DNA and A-RNA both qualitatively as well as quantitatively with an excellent precision. Fig 4b shows the computed current at 1 V for both dsDNA as well as dsRNA for 2500 frames throughout the 100 ns MD simulation trajectory. This graph provides the in-depth understanding of the dynamic variation of electrical properties of the nucleic acid as the MD simulation progresses which is computationally very demanding because of the involved quantum chemical calculations. The order of magnitude of the current is pA which is also in accordance with previously established results[33, 39]. Also, to check whether our approach can reproduce the well-known linear dependence of dsRNA conductance with length in the hopping regime[40], we calculate the electrical conductance of dsRNA of d-(CGCGA$_n$U$_n$CGCG) with n = 2 to 5 (Fig 4c). We find that the conductance decreases with increasing dsRNA length. These results signify the validity and the utility of our ML model for the prediction of electronic couplings.

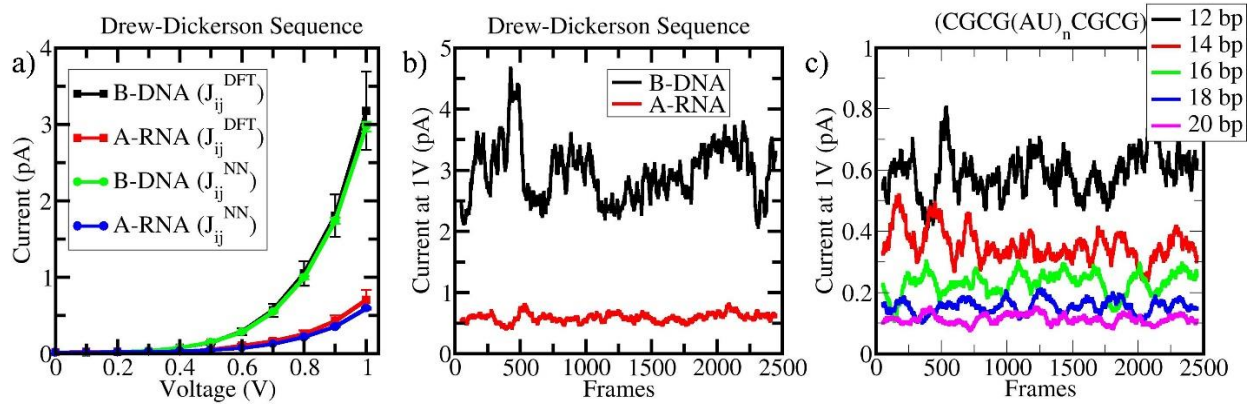

Fig. 4 a) A comparison of I-V characteristics for B-DNA and A-RNA using DFT calculated electronic couplings for 100 morphologies and using NN predicted electronic couplings for 2500 morphologies. b) Variation of current at an applied potential bias of 1 V for B-DNA (d-(CGCGAATTCGCG)) and A-RNA (d-(CGCGAAUUCGCG)) with 2500 MD-simulated snapshots. Clearly, B-DNA shows higher conductance than A-RNA which is in accordance with earlier results[33]. c) Value of current at 1 V for d-(CGCGA$_n$U$_n$CGCG) with n=2 to 5, for 2500 MD-simulated morphologies. Clearly, as the length of the dsRNA increases, the current decreases.

DNA mismatch and repair are fundamental evolutionary-conserved phenomena which play a key role in maintaining the genomic stability[70-72]. Defects in mismatch repair mechanism can lead to highly elevated rates of base substitution and mutations, consequently causing cancer and tumour [70, 71, 73]. DNA-mediated charge transfer phenomenon has been found well-suited for the detection of mismatch in dsDNA sequence[74, 75]. To explore the capabilities of our ML model, we predict the charge transport properties of a dsDNA with a single A-G base pair mismatch as shown in Fig. 5a. Fig. 5b shows the variation of current through the dsDNA with and without a mismatch at an applied potential bias of 1 V. Mismatch of a single base pair in dsDNA can lead to variations in its conductance[76, 77] which are fully captured using our ML approach. The order of magnitude of current is pA which agrees with previous established results[33, 39].

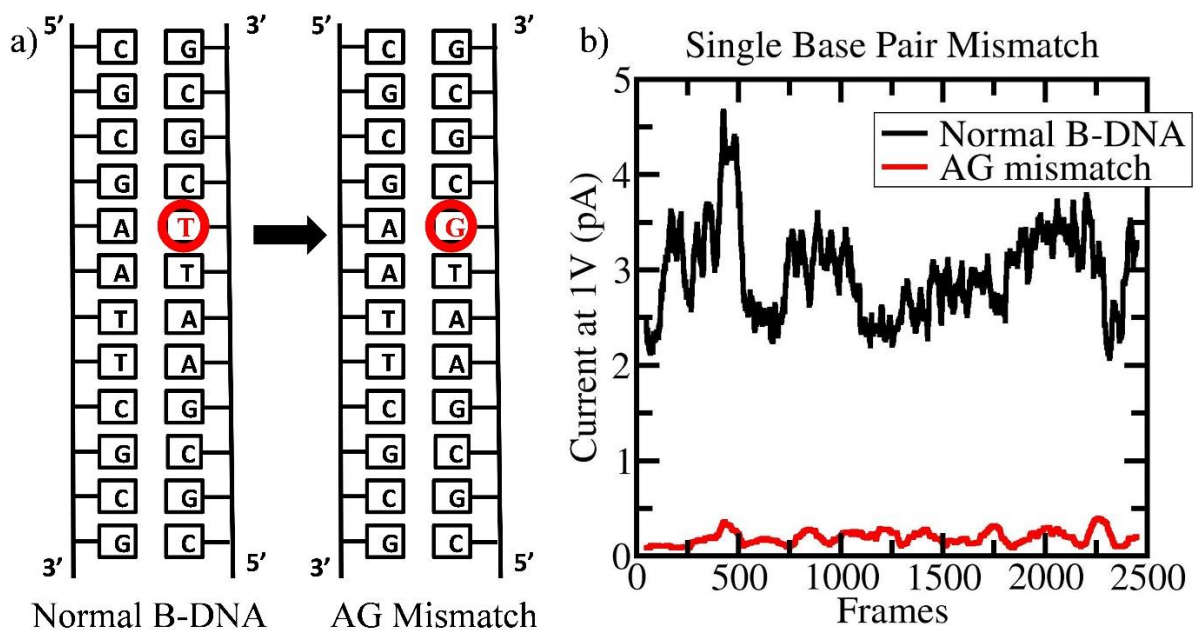

Fig. 5 a) Schematic representation showing a single mismatch in Drew-Dickerson B-DNA sequence. b) The plot of current at an applied potential bias of 1 V for 2500 MD-simulated-morphologies of both Drew-Dickerson sequence with and without base-pair mismatch using the NN predicted electronic couplings shows that a single mismatch can be detected using electrical properties of dsDNA.

## Section 4. Summary and Conclusion

In conclusion, we have prepared an NN model using Coulomb Matrix that can predict the transfer integral between any two dsDNA/dsRNA bases from only the atomic coordinate information. This helps to bypass the most time-consuming DFT calculation part and calculate the dsDNA/dsRNA conductance of any given sequence in orders of magnitude less timescale. Our model captures the dependence of electronic properties of nucleobases on their relative orientation as well as their chemistry. Since this model depends only on the atomic coordinates of the charged entities, it can be easily extended to many other molecular systems such as proteins, dendrimer melts, organic semiconductors, and so on. We showcase the use of our NN model to reproduce well-known charge transport properties of dsDNA and dsRNA systems. As a direct application of our NN model, we show that charge transport properties of DNA can be used to detect even a single base-pair mismatch in dsDNA systems. We believe that our robust ML model to compute the charge transport parameters between any two different entities will significantly advance the field of molecular electronics and DNA nanotechnology.

## Conflicts of interest

There are no conflicts of interest to declare.

## Acknowledgements

We thank DST, India for computational support through TUE-CMS computational facility. A.A. thanks MHRD, India for the generous fellowship. Vinayak acknowledges support from a fellowship provided by the Kishore Vaigyanik Protsahan Yojana (KVPY) scheme of the DST, Government of India.

## References

1. Korol, R. and D. Segal, *Machine Learning Prediction of DNA Charge Transport.* The journal of physical chemistry B, 2019. **123**(13): p. 2801-2811.
2. Carrasquilla, J. and R.G. Melko, *Machine learning phases of matter.* Nature Physics, 2017. **13**(5): p. 431-434.
3. Cleophas, T.J., A.H. Zwinderman, and H.I. Cleophas-Allers, *Machine learning in medicine*. Vol. 9. 2013: Springer.
4. Butler, K.T., et al., *Machine learning for molecular and materials science.* Nature, 2018. **559**(7715): p. 547-555.
5. Réda, C., E. Kaufmann, and A. Delahaye-Duriez, *Machine learning applications in drug development.* Computational and Structural Biotechnology Journal, 2020. **18**: p. 241-252.
6. Friederich, P., et al., *Machine learning of correlated dihedral potentials for atomistic molecular force fields.* Scientific reports, 2018. **8**(1): p. 1-8.
7. Scherer, C., et al., *Kernel-based machine learning for efficient simulations of molecular liquids.* Journal of chemical theory and computation, 2020. **16**(5): p. 3194-3204.
8. Seko, A., A. Togo, and I. Tanaka, *Descriptors for machine learning of materials data*, in *Nanoinformatics*. 2018, Springer, Singapore. p. 3-23.

9.	Welborn, M., L. Cheng, and T.F. Miller III, *Transferability in machine learning for electronic structure via the molecular orbital basis.* Journal of chemical theory and computation, 2018. **14**(9): p. 4772-4779.
10.	Quaranta, V., J.r. Behler, and M. Hellström, *Structure and Dynamics of the Liquid–Water/Zinc-Oxide Interface from Machine Learning Potential Simulations.* The Journal of Physical Chemistry C, 2018. **123**(2): p. 1293-1304.
11.	Fanourgakis, G.S., et al., *A Robust Machine Learning Algorithm for the Prediction of Methane Adsorption in Nanoporous Materials.* The Journal of Physical Chemistry A, 2019. **123**(28): p. 6080-6087.
12.	Wang, C.-I., et al., *Machine Learning for Predicting Electron Transfer Coupling.* The Journal of Physical Chemistry A, 2019. **123**(36): p. 7792-7802.
13.	Cuevas-Zuviría, B. and L.F. Pacios, *Analytical Model of Electron Density and Its Machine Learning Inference.* Journal of Chemical Information and Modeling, 2020. **60**(8): p. 3831-3842.
14.	Pereira, F., et al., *Machine Learning Methods to Predict Density Functional Theory B3LYP Energies of HOMO and LUMO Orbitals.* Journal of Chemical Information and Modeling, 2017. **57**(1): p. 11-21.
15.	Rauer, C. and T. Bereau, *Hydration free energies from kernel-based machine learning: Compound-database bias.* The Journal of chemical physics, 2020. **153**(1): p. 014101.
16.	Kitchin, J.R., *Machine learning in catalysis.* Nature Catalysis, 2018. **1**(4): p. 230-232.
17.	Lederer, J., et al., *Machine Learning–Based Charge Transport Computation for Pentacene.* Advanced Theory and Simulations, 2019. **2**(2): p. 1800136.
18.	Kamal, D., et al., *A charge density prediction model for hydrocarbons using deep neural networks.* Machine Learning: Science and Technology, 2020. **1**(2): p. 025003.
19.	Bleiziffer, P., K. Schaller, and S. Riniker, *Machine Learning of Partial Charges Derived from High-Quality Quantum-Mechanical Calculations.* Journal of Chemical Information and Modeling, 2018. **58**(3): p. 579-590.
20.	Sun, B., M. Fernandez, and A.S. Barnard, *Machine Learning for Silver Nanoparticle Electron Transfer Property Prediction.* Journal of Chemical Information and Modeling, 2017. **57**(10): p. 2413-2423.
21.	Takahashi, A., A. Seko, and I. Tanaka, *Linearized machine-learning interatomic potentials for non-magnetic elemental metals: Limitation of pairwise descriptors and trend of predictive power.* The Journal of chemical physics, 2018. **148**(23): p. 234106.
22.	Guo, Y., et al., *Deep learning for visual understanding: A review.* Neurocomputing, 2016. **187**: p. 27-48.
23.	Batra, R., et al., *Screening of Therapeutic Agents for COVID-19 Using Machine Learning and Ensemble Docking Studies.* The journal of physical chemistry letters, 2020.
24.	Walter, V., et al., *A machine learning study of the two states model for lipid bilayer phase transitions.* Physical Chemistry Chemical Physics, 2020.
25.	Caylak, O., A. Yaman, and B.r. Baumeier, *Evolutionary approach to constructing a deep feedforward neural network for prediction of electronic coupling elements in molecular materials.* Journal of chemical theory and computation, 2019. **15**(3): p. 1777-1784.
26.	Padula, D., J.D. Simpson, and A. Troisi, *Combining electronic and structural features in machine learning models to predict organic solar cells properties.* Materials Horizons, 2019. **6**(2): p. 343-349.
27.	Sahu, H., et al., *Toward predicting efficiency of organic solar cells via machine learning and improved descriptors.* Advanced Energy Materials, 2018. **8**(24): p. 1801032.
28.	Krämer, M., et al., *Charge and Exciton Transfer Simulations using Machine-Learned Hamiltonians.* Journal of chemical theory and computation, 2020.

29. Bag, S., A. Aggarwal, and P.K. Maiti, *Machine Learning Prediction of Electronic Coupling between the Guanine Bases of DNA.* The Journal of Physical Chemistry A, 2020.
30. Boon, E.M. and J.K. Barton, *Charge transport in DNA.* Current opinion in structural biology, 2002. **12**(3): p. 320-329.
31. Merino, E.J., A.K. Boal, and J.K. Barton, *Biological contexts for DNA charge transport chemistry.* Current opinion in chemical biology, 2008. **12**(2): p. 229-237.
32. O'brien, E., et al., *The [4Fe4S] cluster of human DNA primase functions as a redox switch using DNA charge transport.* Science, 2017. **355**(6327): p. eaag1789.
33. Aggarwal, A., et al., *Multiscale modelling reveals higher charge transport efficiencies of DNA relative to RNA independent of mechanism.* Nanoscale, 2020.
34. Aggarwal, A., et al., *What do we know about DNA mechanics so far?* Current opinion in structural biology, 2020. **64**: p. 42-50.
35. Zhuravel, R., et al., *Backbone charge transport in double-stranded DNA.* Nature nanotechnology, 2020: p. 1-5.
36. Bartels, P.L., et al., *Electrochemistry of the [4Fe4S] cluster in base excision repair proteins: tuning the redox potential with DNA.* Langmuir, 2017. **33**(10): p. 2523-2530.
37. Qi, J., et al., *Unified model for conductance through DNA with the Landauer-Büttiker formalism.* Physical Review B, 2013. **87**(8): p. 085404.
38. Genereux, J.C. and J.K. Barton, *Mechanisms for DNA charge transport.* Chemical reviews, 2009. **110**(3): p. 1642-1662.
39. Aggarwal, A., S. Bag, and P.K. Maiti, *Remarkable similarity of force induced dsRNA conformational changes to stretched dsDNA and their detection using electrical measurements.* Physical Chemistry Chemical Physics, 2018. **20**(45): p. 28920-28928.
40. Bag, S., et al., *Dramatic changes in DNA conductance with stretching: structural polymorphism at a critical extension.* Nanoscale, 2016. **8**(35): p. 16044-16052.
41. Xu, et al., *Direct Conductance Measurement of Single DNA Molecules in Aqueous Solution.* Nano letters, 2004. **4**(6): p. 1105-1108.
42. Li, Y., et al., *Detection and identification of genetic material via single-molecule conductance.* Nature nanotechnology, 2018. **13**(12): p. 1167.
43. Artés, J.M., et al., *Conformational gating of DNA conductance.* Nature communications, 2015. **6**: p. 8870.
44. Bag, S. and P.K. Maiti, *Tuning molecular fluctuation to boost the conductance in DNA based molecular wires.* Physical Chemistry Chemical Physics, 2019. **21**(42): p. 23514-23520.
45. Troisi, A., A. Nitzan, and M.A. Ratner, *A rate constant expression for charge transfer through fluctuating bridges.* The Journal of chemical physics, 2003. **119**(12): p. 5782-5788.
46. Troisi, A. and G. Orlandi, *Hole migration in DNA: a theoretical analysis of the role of structural fluctuations.* The journal of physical chemistry B, 2002. **106**(8): p. 2093-2101.
47. Wolter, M., et al., *Microsecond Simulation of Electron Transfer in DNA: Bottom-Up Parametrization of an Efficient Electron Transfer Model Based on Atomistic Details.* The journal of physical chemistry B, 2017. **121**(3): p. 529-549.
48. Kubař, T. and M. Elstner, *Efficient algorithms for the simulation of non-adiabatic electron transfer in complex molecular systems: application to DNA.* Physical Chemistry Chemical Physics, 2013. **15**(16): p. 5794-5813.
49. Woiczikowski, P.B., et al., *Combined density functional theory and Landauer approach for hole transfer in DNA along classical molecular dynamics trajectories.* The Journal of chemical physics, 2009. **130**(21): p. 06B608.
50. Wolter, M., M. Elstner, and T. Kubař, *Charge transport in desolvated DNA.* The Journal of chemical physics, 2013. **139**(12): p. 09B648_1.

51. Song, B., M. Elstner, and G. Cuniberti, *Anomalous conductance response of DNA wires under stretching.* Nano letters, 2008. **8**(10): p. 3217-3220.
52. Venkatramani, R., et al., *Nucleic acid charge transfer: black, white and gray.* Coordination chemistry reviews, 2011. **255**(7-8): p. 635-648.
53. Case, D., et al., *Duke, TJ Giese, H.* Gohlke, AW Goetz, N. Homeyer, S. Izadi, P. Janowski, J. Kaus, A. Kovalenko, TS Lee, S. LeGrand, P. Li, T. Luchko, R. Luo, B. Madej, KM Merz, G. Monard, P. Needham, H. Nguyen, HT Nguyen, I. Omelyan, A. Onufriev, DR Roe, A. Roitberg, R. Salomon-Ferrer, CL Simmerling, W. Smith, J. Swails, RC Walker, J. Wang, RM Wolf, X. Wu, DM York and PA Kollman Amber, 2015.
54. Jorgensen, W.L., et al., *Comparison of simple potential functions for simulating liquid water.* The Journal of chemical physics, 1983. **79**(2): p. 926-935.
55. Pérez, A., et al., *Refinement of the AMBER force field for nucleic acids: improving the description of α/γ conformers.* Biophysical journal, 2007. **92**(11): p. 3817-3829.
56. Zgarbová, M., et al., *Refinement of the Cornell et al. nucleic acids force field based on reference quantum chemical calculations of glycosidic torsion profiles.* Journal of chemical theory and computation, 2011. **7**(9): p. 2886-2902.
57. Zgarbová, M., et al., *Toward improved description of DNA backbone: revisiting epsilon and zeta torsion force field parameters.* Journal of chemical theory and computation, 2013. **9**(5): p. 2339-2354.
58. Zgarbová, M., et al., *Refinement of the Sugar–Phosphate backbone torsion beta for AMBER force fields improves the description of Z-and B-DNA.* Journal of chemical theory and computation, 2015. **11**(12): p. 5723-5736.
59. Sahoo, A.K., B. Bagchi, and P.K. Maiti, *Understanding enhanced mechanical stability of DNA in the presence of intercalated anticancer drug: Implications for DNA associated processes.* The Journal of chemical physics, 2019. **151**(16): p. 164902.
60. Naskar, S., et al., *Tuning the Stability of DNA Nanotubes with Salt.* The Journal of Physical Chemistry C, 2019. **123**(14): p. 9461-9470.
61. Humphrey, W., A. Dalke, and K. Schulten, *VMD: visual molecular dynamics.* Journal of molecular graphics, 1996. **14**(1): p. 33-38.
62. Su, S., et al., *Mispair specificity of methyl-directed DNA mismatch correction in vitro.* Journal of Biological Chemistry, 1988. **263**(14): p. 6829-6835.
63. Marcus, R.A., *Electron transfer reactions in chemistry. Theory and experiment.* Reviews of Modern Physics, 1993. **65**(3): p. 599.
64. Endres, R.G., D.L. Cox, and R.R. Singh, *Colloquium: The quest for high-conductance DNA.* Reviews of Modern Physics, 2004. **76**(1): p. 195.
65. Frisch, M., et al., *Gaussian∼09 Revision D. 01.* 2014.
66. Rühle, V., et al., *Microscopic simulations of charge transport in disordered organic semiconductors.* Journal of chemical theory and computation, 2011. **7**(10): p. 3335-3345.
67. Hansen, K., et al., *Assessment and validation of machine learning methods for predicting molecular atomization energies.* Journal of chemical theory and computation, 2013. **9**(8): p. 3404-3419.
68. Rupp, M., et al., *Fast and accurate modeling of molecular atomization energies with machine learning.* Physical review letters, 2012. **108**(5): p. 058301.
69. Pedregosa, F., et al., *Scikit-learn: Machine learning in Python.* the Journal of machine Learning research, 2011. **12**: p. 2825-2830.
70. Iyer, R.R., et al., *DNA mismatch repair: functions and mechanisms.* Chemical reviews, 2006. **106**(2): p. 302-323.

71. Li, G.-M., *Mechanisms and functions of DNA mismatch repair.* Cell research, 2008. **18**(1): p. 85-98.
72. Philip, A. and H. James, *Detection of a single DNA base-pair mismatch using an anthracene-tagged fluorescent probe.* Chemical Communications, 2006(48): p. 5003-5005.
73. De la Chapelle, A., *Genetic predisposition to colorectal cancer.* Nature Reviews Cancer, 2004. **4**(10): p. 769-780.
74. Drummond, T.G., M.G. Hill, and J.K. Barton, *Electrochemical DNA sensors.* Nature biotechnology, 2003. **21**(10): p. 1192-1199.
75. Kelley, S.O., et al., *Single-base mismatch detection based on charge transduction through DNA.* Nucleic acids research, 1999. **27**(24): p. 4830-4837.
76. Guo, X., et al., *Conductivity of a single DNA duplex bridging a carbon nanotube gap.* Nature nanotechnology, 2008. **3**(3): p. 163.
77. Starikov, E., et al., *Effects of molecular motion on charge transfer/transport through DNA duplexes with and without base pair mismatch.* Molecular Simulation, 2006. **32**(9): p. 759-764.

# Predicting the DNA Conductance using Deep Feed Forward Neural Network Model: Supplementary information

Abhishek Aggarwal[1, a], Vinayak[2, a], Saientan Bag[1], Chiranjib Bhattacharyya[3], Umesh V. Waghmare[4] and Prabal K Maiti[1, *]

[1]Center for Condensed Matter Theory, Department of Physics, Indian Institute of Science, Bangalore 560012

[2]Undergraduate program, Indian Institute of Science, Bangalore 560012

[3]Department of Computer Science and Automation, Indian Institute of Science, Bangalore 560012

[4]Theoretical Sciences Unit, Jawaharlal Nehru Center for Advanced Scientific Research, Jakkur P.O., Bangalore 560064, India

[a] Equal Contribution
*email: maiti@iisc.ac.in, phone: 091-80-2293-2865

## Section S1. Grid search for hyperparameters

Using the training dataset (which forms 0.8 partition of the total dataset), we do a k-fold cross validation test, with k=5, to find the best set of hyperparameters for our FFNN. The search is done among the set shown in

Table **S1**.

| Hyperparameter | Grid search values |
|---|---|
| Activation function | Identity, **logistic,** tanh, ReLU |
| Nodes and layers | (100,80,10), (**100,50,10**), (100,20,10), (80,20) |
| Alpha (coefficient of L2 regularizer) | 0.1, 0.01, 0.001, **0.0001**, 0.00001 |
| Solver | Constant, invscaling, **adaptive** |
| Learning rate | **Adam**, sgd |

**Table S1:** Hyperparameters among which the grid search is done. The bold-faced hyperparameters are used for further calculations in this work.

## Section S2. DFT calculated electronic coupling vs NN predicted electronic coupling plots for dsDNA/dsRNA base pairs

In the main article, we present the validation and test sets for base pairs AA, AT, CC, CG, GG and TT along with their MAE. Here we plot the remaining base pairs, i.e., GU, CU, UU, AG, AC, and AU in figure S1.

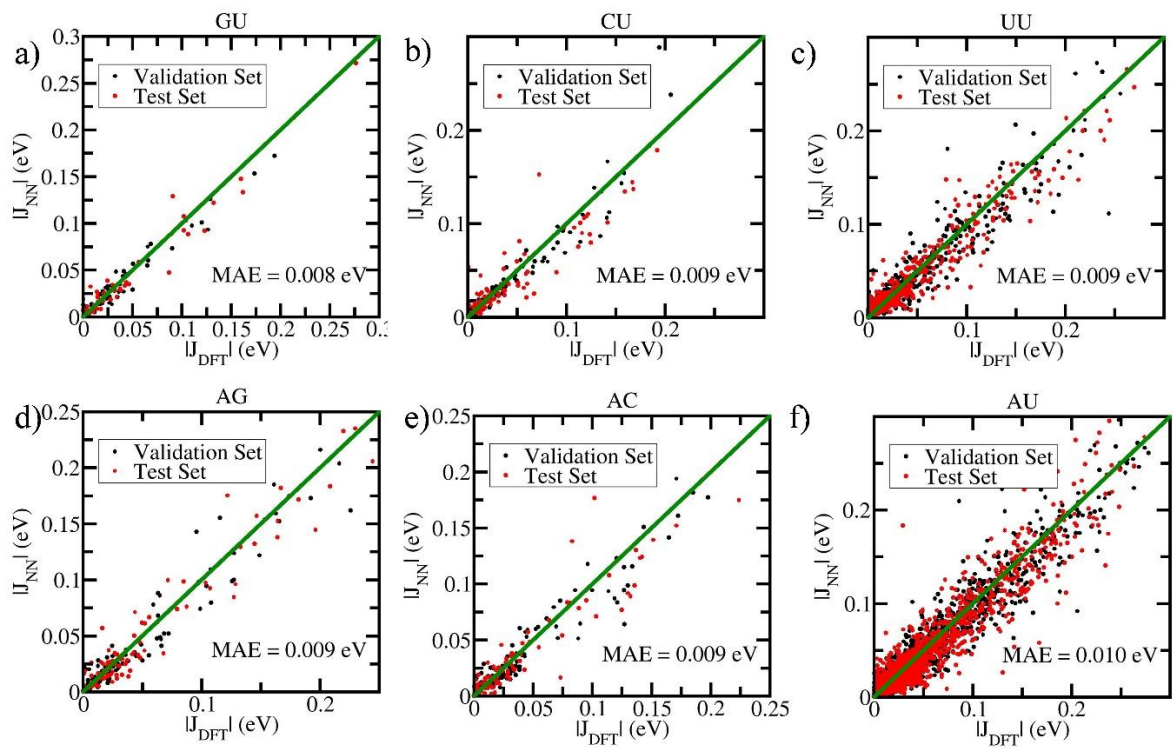

**Figure S2** NN prediction results for different base-pair combinations. Correlation between the NN predicted values of the electronic coupling vs the DFT calculated electronic couplings for a) GU, b) CU, c) UU, d) AG, e) AC and f) AU base-pairs. The MAE for all the base pairs is less than 0.010 eV. The green lines represent the ideal case of NN predicted electronic couplings to be equal with DFT calculated values.